\documentclass{elsart}
\usepackage{english}
\usepackage{epsfig}
\usepackage{amsfonts}
\newcommand\nn{\nonumber\\}

\newcommand{\be}{\begin{equation}}
\newcommand{\ee}{\end{equation}}
\newcommand{\ba}{\begin{eqnarray}}
\newcommand{\ea}{\end{eqnarray}}
\newcommand{\msbar}{\overline{\mbox{\rm MS}}}
\newcommand{\eq}{Eq.~}
\newcommand{\I}{{\rm i}}
\newcommand{\x}{{\bf x}}
\newcommand{\Ocal}{{\mathcal O}}
\newcommand{\D}{{\mathcal D}}
\newcommand{\rmi}[1]{{\mbox{\scriptsize #1}}}
\begin{document}
\begin{frontmatter}
\begin{flushright}
CERN-TH/98-295\\
HIP-1998-60/TH\\
hep-ph/9809334
\end{flushright}

\vspace*{15.3cm}

\begin{flushleft}
CERN-TH/98-295\\
HIP-1998-60/TH\\
September 1998
\end{flushleft}

\vspace*{-17cm}

\title{Vortex tension as an order parameter in three-dimensional 
U(1)+Higgs theory}

\author[CERN,TFO]{K.~Kajantie\thanksref{keijo}},
\author[CERN,TFO]{M.~Laine\thanksref{mikko}},
\author[HIP,Biel]{T.~Neuhaus\thanksref{thomas}},
\author[Swan]{J.~Peisa\thanksref{janne}},
\author[TFO,HIP]{A.~Rajantie\thanksref{arttu}} and
\author[Nord]{K.~Rummukainen\thanksref{kari}}

\address[CERN]{Theory Division, CERN, CH-1211 Geneva 23, Switzerland}
\address[TFO]{Dept.\ of Physics, P.O. Box 9,
FIN-00014 Univ.\ of Helsinki, Finland}
\address[HIP]{Helsinki Institute of Physics, P.O. Box 9,
FIN-00014 Univ.\ of Helsinki, Finland}
\address[Biel]{\mbox{Fakult\"at f\"ur Physik, Univ.\ Bielefeld,
P.O. Box 100131, D-33501 Bielefeld, FRG}}
\address[Swan]{\mbox{Dept.\ of Physics, Univ.\ of Wales Swansea,
Singleton Park, Swansea SA2 8PP, UK}}
\address[Nord]{Nordita, Blegdamsvej 17, DK-2100 Copenhagen \O, Denmark}

\thanks[keijo]{keijo.kajantie@cern.ch}
\thanks[mikko]{mikko.laine@cern.ch}
\thanks[thomas]{thomas.neuhaus@helsinki.fi}
\thanks[janne]{pyjanne@swansea.ac.uk}
\thanks[arttu]{arttu.rajantie@helsinki.fi}
\thanks[kari]{kari@nordita.dk}




\vspace*{-0.5cm}

\abstract
We use lattice Monte Carlo simulations to study non-perturbatively the 
tension, i.e.~the free energy per unit length, of an infinitely long vortex 
in the three-dimen\-sio\-nal U(1)+Higgs theory. This theory is the low-energy 
effective theory of high-temperature scalar electrodynamics, the standard
framework for cosmic string studies. The vortex tension is measured
as a function of the mass parameter at a large value of the Higgs
self-coupling, where the transition between the phases
is continuous. It is shown that the tension gives an order parameter 
that can distinguish between the two phases of the system.
We argue that the vortex tension can describe the physics of long strings
without lattice artifacts, unlike vortex network percolation.
\endabstract
\end{frontmatter}

\newpage

\section{Introduction}
During the early stages of its evolution, the Universe went through
a series of phase transitions. Topological defects, such as
cosmic strings, domain walls and magnetic monopoles, may have been 
created as a result of these transitions~\cite{ref:defreviews}.
It is possible that
the topological defects have observable consequences even 
in the present Universe.

It is customary to view the topological defects as classical solutions
obtained from the high-temperature effective potential of the
system. However, the validity of this mean-field approach is uncertain,
especially in gauge theories. Indeed, there is no
symmetry breaking related to the Higgs mechanism \cite{ref:elitzur},
and in some cases the phases are analytically
connected to each other \cite{ref:fradkinshenker}. 
Clearly the mean-field approximation does
not work near the transition in such cases. It is therefore necessary
to investigate how the inclusion of the full effect of fluctuations
changes the picture of defect formation.

To study the properties of string-like defects in gauge theories,
it is most convenient to concentrate on the Nielsen-Olesen vortices
\cite{ref:nielsenolesen}
of the U(1) gauge+Higgs theory. Although this is much simpler than
realistic theories containing cosmic strings, 
it permits a first-principles field theoretic study.
In Ref.~\cite{ref:dimred} 
an effective three-dimensional (3d) theory was constructed to describe 
the thermodynamical behavior of the full four-dimensional (4d) theory
at high temperatures near the phase transition.
The effective theory, the 3d U(1)+Higgs theory, is equivalent
to the Ginzburg-Landau theory of superconductivity \cite{ref:kleinert}.
It contains fewer degrees of freedom than the original theory and
its ultraviolet behavior is also better understood
\cite{ref:contlatt}, which makes it
much more suitable for numerical simulations.

In discussing phase transitions, the first question is to find a proper
variable to signal one. Although, in a gauge theory, there is no local 
gauge-invariant order parameter, the first order transition 
in the U(1)+Higgs theory can be localized
by finding a discontinuity in, say, $|\phi|^2$ 
\cite{ref:dfk,ref:u1big}. 
This does not help if the transition is continuous.
Then a non-local order parameter may exist. For U(1)+Higgs one such
is the photon mass \cite{ref:u1big} which vanishes identically in
the symmetric (Coulomb) phase and is non-zero in the broken (Higgs)
phase.

For the U(1)+Higgs theory one may suggest further order parameters
related to the existence of vortices. In fact, several different
effective models have been proposed (see, e.g., 
\cite{ref:kleinert,ref:kovner}).
These cannot replace a first-principles numerical study. In
Ref.~\cite{ref:vortex} the total density of thermal vortex-loop
excitations was studied as an order parameter.
However, this suffers from the difficulty of performing a continuum
limit: small vortices also appear as lattice artifacts.

In scalar theories, it has been proposed that one should instead
look at the density of {\em infinitely long} vortices, i.e., 
percolation \cite{ref:perco} (see also~\cite{ref:cgis}). 
However, as is well-known, 
this definition could be subject to lattice 
artifacts (for a review, see~\cite{ref:isi}). The basic problem
is that ``percolation'' is a geometrical property, 
not directly related to the free energy of the system,
and possibly sensitive to how the discretization 
needed for simulations is made.

In this paper, we suggest a better alternative in the context of gauge
theories:
the tension of an infinitely long vortex, i.e., its free 
energy per unit length. We show how one puts a vortex on a lattice
and how one measures its free energy
using Monte Carlo simulations. We show that it really is an order
parameter, vanishing in the $V\to\infty$ limit in the same phase as 
the photon mass. The result
is given in continuum $\msbar$ units, which makes it
possible to compare it with the mean-field value \cite{ref:jacobsrebbi}. 
The vortex tension thus defined is a physical observable and 
has a well-defined continuum limit. We suggest that 
the vortex tension is a good first-principles tool 
for discussing the physics of long vortices, and might
perhaps also be used as an input in some phenomenological
non-equilibrium estimates in the regime where
mean-field theory does not work.

The structure of the paper is the following. 
The U(1)+Higgs theory is defined and its properties are reviewed
in Sec.~\ref{sect:abhiggs}.
In Sec.~\ref{sect:insvort} we describe how the vortex tension can be 
defined and measured, and in Sec.~\ref{sect:relation} we
relate it to the phase structure of the theory.
The details of the simulations, as well as their results, are given
in Sec.~\ref{sect:simulations}. Sec.~\ref{sect:conclusions}
contains the conclusions.

\section{The U(1)+Higgs theory}
\label{sect:abhiggs}
In Ref.~\cite{ref:dimred}, the heavy degrees of freedom were integrated
out from the 4d high-temperature U(1)+Higgs theory 
(see also~\cite{ref:u1big,ref:joa}). The
resulting 3d theory has the action
\be
\label{equ:contact}
S=\int\d^3x\left[
\frac{1}{4}F_{ij}F_{ij}
+\left|\left(\partial_i-\I
A_i\right)\phi\right|^2
+y\left|\phi\right|^2
+x\left|\phi\right|^4
\right],
\ee
where $F_{ij}=\partial_iA_j-\partial_jA_i$. Everything is expressed as
dimensionless quantities, by scaling with appropriate powers of
$e_3^2$, and the mass parameter $y$ is
renormalized in the $\msbar$ scheme with the scale $\mu=1$.

\begin{figure}
\hspace*{1cm}\epsfig{file=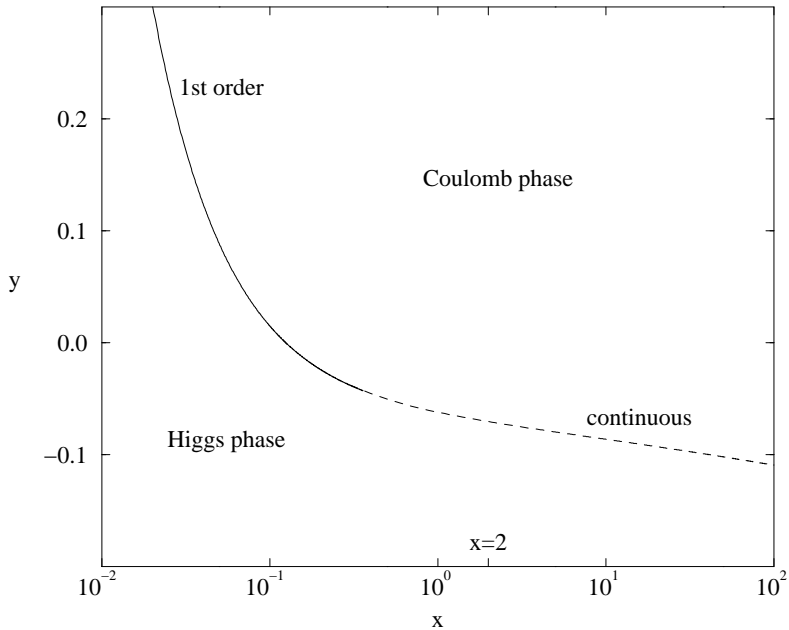,width=12cm}
\caption{
\label{fig:phases}
The phase diagram of the 3d U(1)+Higgs theory. The vortex tension 
measurements in this paper are carried out in the region of
continuous transitions ($x=2$) for $-1.6\le y\le 0.5$.}
\end{figure}

The phase diagram of the theory (see Fig.~\ref{fig:phases}) has been studied 
with lattice simulations in Refs.~\cite{ref:dfk,ref:u1big}.
On the lattice, there are two alternative formulations, the compact
and the non-compact one, but they are expected to have the same
continuum limit in 3d.
It was shown in Ref.~\cite{ref:borgsnill} that
in the non-compact lattice formulation there are two phases, 
the Higgs and the Coulomb phase.
There is no symmetry breaking and no local order parameter, but
the mass of the photon acts as a non-local 
order parameter: it is zero in the Coulomb
phase and non-zero in the Higgs phase.
This differs from the compact lattice formulation 
at any finite lattice spacing, 
and from the SU(2)+Higgs model, in which cases the phases are 
analytically connected \cite{ref:fradkinshenker,ref:su2}.

At the value $x=2$ that we will use in this paper, the transition
between the two phases has been 
observed to be continuous, but quite different from
standard symmetry breaking transitions \cite{ref:u1big}.
At small $x$,
a first-order transition was found in accordance with
perturbation theory \cite{ref:colemanweinberg}.

The field equations derived from the action 
in \eq(\ref{equ:contact})
have Nielsen-Olesen vortices as solutions when $y<0$ 
\cite{ref:nielsenolesen}. The vortex tension 
can be calculated in the mean-field approximation by solving the
field equations numerically. The result can be written in the form
\be
\label{equ:classT}
T_{\rm MF}=\frac{\Delta S}{L}=-\frac{y}{x}\pi{\mathcal E}(\sqrt{2x}),
\ee
where the function ${\mathcal E}$,
with the value ${\mathcal E}(1)=1$, has been calculated
numerically in, e.g., Ref.~\cite{ref:jacobsrebbi}. 
From there we interpolate that
${\mathcal E}(2)\approx 1.32$, which
value we use for comparison with our lattice results.

To study the tension $T$
non-perturbatively in the quantum theory,
we discretize the space and define the system on a lattice with lattice
spacing~$a$.
We use the non-compact formulation, i.e.~the link field is a real number
instead of a phase angle.
The lattice action corresponding to the continuum theory 
in \eq(\ref{equ:contact}) is
\ba
S & = & \beta_G \!\!\sum_{\x,i<j}
\frac{1}{2}\alpha_{ij}^2(\x)
  -{2\over\beta_G} \sum_{\x, i} {\mbox{Re}}\, 
\phi^*(\x) U_i(\x)\phi(\x+\hat{\imath})\nonumber\\   
  & & +\left[...\right]\sum_{\x} \phi^*(\x) \phi(\x)
+{x\over\beta_G^3}\sum_{\x} \left[\phi^*(\x)\phi(\x)\right]^2, 
\label{equ:lattact} 
\ea
where 
$\alpha_{ij}(\x)=\alpha_i(\x)+\alpha_j(\x+\hat\imath)-
\alpha_i(\x+\hat\jmath)-\alpha_j(\x)$, $U_i(\x)=\exp[i\alpha_i(\x)]$,
$\x=(x_1,x_2,x_3)$ with $1\le x_i\le N_i$,
\be
 \beta_G={1\over a},
\ee
and the coefficient of the quadratic term is~\cite{ref:contlatt}
\ba
\left[...\right]&=&{1\over\beta_G}\left[6+{y\over\beta_G^2}
-{3.1759115(1+2x)\over2\pi\beta_G}\right.\nn
&&\left.-\frac{
(-4+8x-8x^2)(\log6\beta_G+0.09)-1.1+4.6x}{16\pi^2\beta_G^2}\right].
\ea
Note how remarkably simple and analytic the lattice-continuum relation is.

To give a gauge-invariant definition for a vortex on a lattice,
we define for each link~\cite{ref:vortex,ref:old} (We choose the opposite 
sign here!)
\be
\label{equ:link}
Y_{(\x,\x+\hat\imath)}=\alpha_i(\x)-
[\alpha_i(\x)+\gamma(\x+\hat\imath)-
\gamma(\x)]_\pi,
\ee
where $\gamma=\arg\phi$ and $[X]_\pi=X+2\pi n_X$ such that 
$[X]_\pi\in(-\pi,\pi]$.
For links in the opposite direction, 
$Y_{(\x,\x-\hat\imath)}=-Y_{(\x-\hat\imath,\x)}.$
For each closed curve $C$ of links on a lattice, we then define
the winding number $n_C$ as
\be
\label{equ:winding}
n_C=\frac{1}{2\pi}\sum_{l\in C}Y_l\in {\mathbb Z}.
\ee
In each configuration, the winding number gives the number of vortices 
passing through the curve $C$. Note that we count vortices with multiple
winding as separate vortices on top of each other.

\begin{figure}

\vspace*{0.5cm}

\begin{center}
\epsfig{file=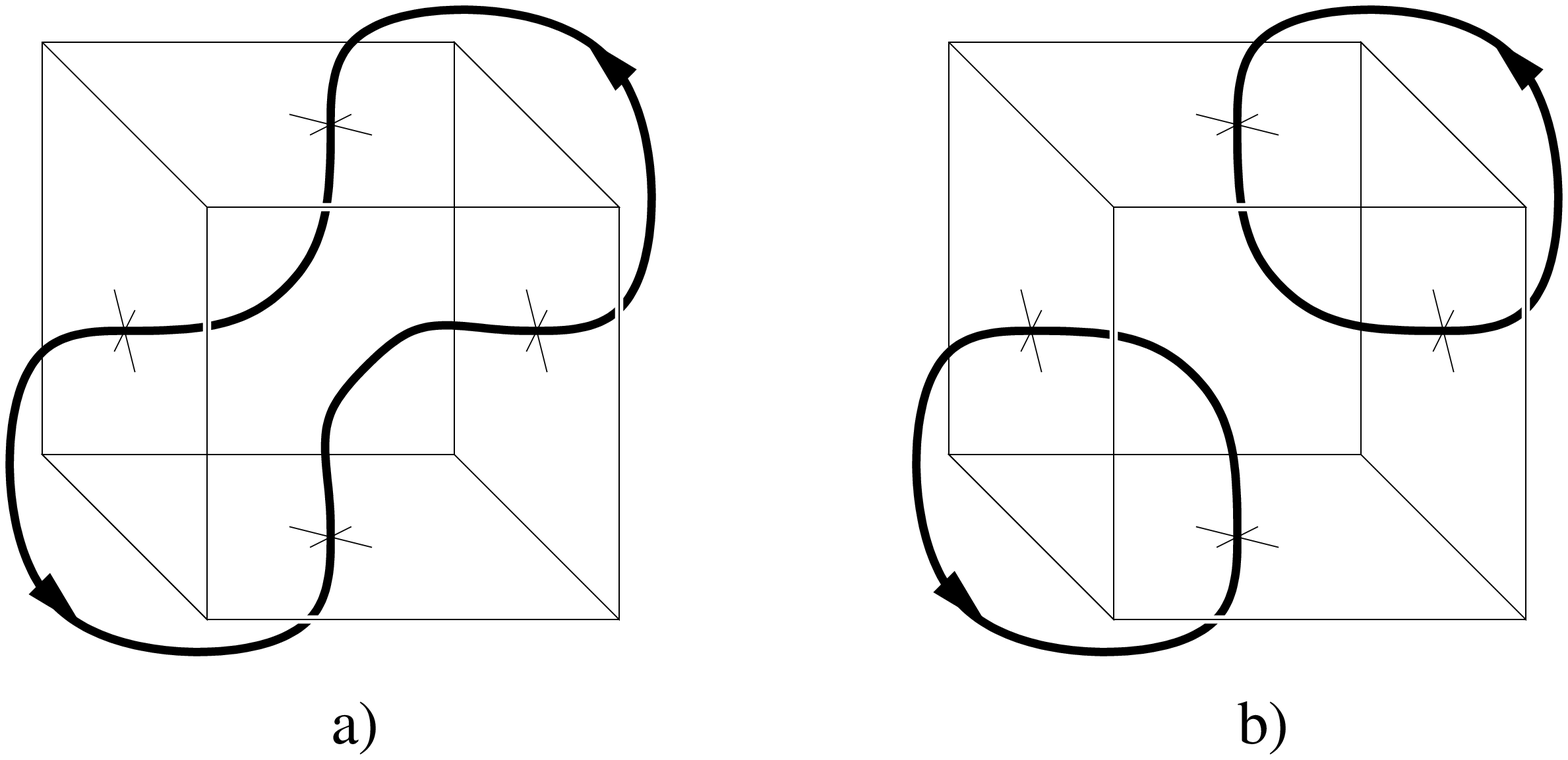,width=10cm}
\end{center}
\caption{
\label{fig:connect}
If two vortices pass through one cell, the vortex tracing algorithm must
decide how to connect the vortices, and this leads to an ambiguity
in the length distribution of vortices.}
\end{figure}

Recently, there have been many attempts to interpret the phase transition
in this and in related models as a percolation transition of vortices 
\cite{ref:perco,ref:cgis}. 
The percolation point $y_p$ is defined as the value of $y$
above which one can in a typical configuration find a vortex that
extends through the whole lattice.
However, there are ambiguities in this procedure if two or more vortices 
meet in one cell (see Fig.~\ref{fig:connect}),
and it is not obvious that
$y_p$ should coincide with $y_c$, which is the transition point
signalled by the non-analytic behavior
in the free energy. For example, in the Ising model, these two transitions 
agree at $d=2$ 
\cite{ref:coniglio}, but $\beta_c\approx 0.95 \beta_p$ at $d>2$ 
\cite{ref:muller}.

The idea behind the percolation studies is that as the transition is
approached from below, the fluctuations can create larger and larger 
vortices, and at the transition point, the typical size of a vortex
loop diverges. In other words, the free energy per unit length $T$ of
a vortex decreases to zero. To avoid the problems of the
percolation approach, we have chosen here to measure $T$ directly.
Since our definition of $T$ is based on the properties of
the gauge field, this approach cannot be used in globally
symmetric theories.

\section{Inserting a vortex and measuring its free energy}
\label{sect:insvort}
Free energies cannot be measured by Monte Carlo simulations, but changes of
free energies can. To measure the free energy of a single
vortex, we define a continuous vortex number (=winding number) $m$,
measure the change of the free energy when $m\to m+dm$, $0<m<1$, and
integrate these changes of $m$ from 0 to~1. The measurements
can as well be done for $m$ varying from 0 to any integer. 

On the lattice, 
we call a field periodic, if it has the same value at
$x_i=N_i+1$ as at $x_i=1$. 
If we require the fields $\phi$ and $\alpha$ to be periodic,
i.e.~use periodic boundary conditions,
then for any planar cross section (say, in the ($x,y)$-plane),
the total flux $\Phi=\sum_\rmi{plane} \alpha_{12}$ 
and the winding $n_{\rm tot}$ defined in \eq(\ref{equ:winding}), 
vanish in every configuration.
The expectation value of an observable 
$\Ocal[\alpha,\phi]$ is then given by
\be
\label{equ:pathint}
\langle \Ocal\rangle=\int_{\rm periodic}\D\alpha\D\phi 
\Ocal[\alpha,\phi]\e^{-S[\alpha,\phi]},
\ee
with $S[\alpha,\phi]$ given in Eq.~(\ref{equ:lattact}).

We now define a non-periodic constant field $\tilde\alpha$ such that 
$\tilde\alpha_2(N_1+1,1,x_3)=2\pi$, 
and 
$\tilde\alpha_i(\x)=0$ otherwise. The magnetic flux through the system in
the $z$-direction, calculated from the background $\tilde\alpha$, is 
$\tilde\Phi=\sum_\rmi{boundary}\tilde\alpha_i=2\pi$.
Note that there is no analogue to this in the compact formulation,
where $\alpha$ is only defined modulo $2\pi$.

Let us now replace $\alpha$ by $\alpha+m\tilde\alpha$ ($m\in\mathbb Z$) 
in the 
integrand of Eq.~(\ref{equ:pathint}):
\be
\label{equ:pathint2}
\langle \Ocal\rangle_m=\int_{\rm periodic}\D\alpha\D\phi 
\Ocal[\alpha+m\tilde\alpha,\phi]\e^{-S[\alpha+m\tilde\alpha,\phi]}.
\ee
Note that a change of variables $\alpha+m\tilde\alpha\rightarrow\alpha$ 
transforms
this to the form of Eq.~(\ref{equ:pathint}) with non-periodic
boundary conditions.
Clearly, all the configurations in this integral have the total flux 
$\Phi=2\pi m$,
and from Eqs.~(\ref{equ:link}), (\ref{equ:winding}) we notice
that the total winding is
$n_{\rm tot}=m$,
i.e.,\ the number of
vortices going through the lattice is $m$. 
It can be shown that any configuration in which $\Phi=2\pi m$ and
$n_{\rm tot}=m$ and the physical quantities $\phi^*\phi$,
$\phi^*(x)U_i(x)\phi(x+\hat\imath)$ and $\alpha_{ij}$ are periodic,
can be gauge transformed to one that is
included in the integral in \eq(\ref{equ:pathint2}).
Therefore, Eq.~(\ref{equ:pathint2}) describes a periodic system in which
the flux and the winding number have been fixed%
\footnote{There are also periodic configurations in which
$\Phi\neq 2\pi n_{\rm tot}$, but their energy is assumed to
diverge logarithmically with the area. Thus we neglect them here.}.

Since $\tilde U_i(\x)=1$ always, and 
$\tilde\alpha_{ij}(\x)=0$ everywhere except 
$\tilde\alpha_{12}(N_1,1,x_3)=2\pi$, we see from Eq.~(\ref{equ:lattact})
that
\be
\label{equ:actchange}
S[\alpha+m\tilde\alpha,\phi]=S[\alpha,\phi]
+\beta_G\sum_{x_3}(2\pi m\alpha_{12}(N_1,1,x_3)+2\pi^2 m^2).
\ee
From this expression it appears as if we had effectively inserted $m$ 
vortices
along the line $(N_1,1,x_3)$, $1\le x_3\le N_3$. However, 
in Eq.~(\ref{equ:pathint2})
we integrate over all periodic field configurations, and thus all
observables are translationally invariant. In fact,
the precise form of $\tilde\alpha$ is irrelevant
as long as its values are multiples of $2\pi$:
only the total flux $\tilde\Phi$ affects the results.

\eq(\ref{equ:pathint2}) means that the free energy of a system 
with $m$ vortices can be written as
\be
\label{equ:freeen}
F_m=-\ln \int_{\rm periodic}\D\alpha\D\phi 
\e^{-S[\alpha+m\tilde\alpha,\phi]}.
\ee
Only when $m \in \mathbb Z$, are the action and the observables 
periodic, and boundary effects are avoided.
This is the well-known flux quantization condition.
Still, $F_m$ is mathematically well defined for any 
$m\in\mathbb R$, and
the tension of the vortex can be written as (cf.~Eq.~(\ref{equ:classT}))
\be
\label{equ:tension}
T=\frac{F_1-F_0}{aN_3}=\frac{1}{aN_3}\int_0^1\d m\frac{\d F_m}{\d m}
=\frac{1}{aN_3}\int_0^1\d m\langle 
\frac{\d S[\alpha+m\tilde\alpha,\phi]}{\d m}\rangle_m.
\ee
Substituting Eq.~(\ref{equ:actchange}) to Eq.~(\ref{equ:tension}) yields
\ba
\label{equ:double_of_m}
T&=&2\pi^2 \beta_G^2 \int_0^1\d m W(m) \nn
&&\equiv 2\pi^2 \beta_G^2 \int_0^1\d m
\left[{1 \over {\pi N_3}} \langle \sum_{x_3} \alpha_{12}(N_1,1,x_3)
\rangle_m+2m
\right].
\ea
It is the quantity $W(m)$ that we calculate with Monte Carlo simulations.


It is useful to inspect what kind of finite size effects one can
expect for $T$.
In the Coulomb phase, the massless photon gives rise to power-like 
finite size effects. Their magnitude can be estimated by considering
the pure gauge theory, since the flux is expected to be distributed
homogeneously in the whole system. In the absence of scalar fields, 
the flux quantization condition 
$m\in\mathbb Z$ does not apply, and the system is translationally invariant
at all $m\in\mathbb R$. Thus
\be
\langle \alpha_{12}(N_1,1,x_3)\rangle_m=
\frac{2\pi m}{N_1N_2}-m\tilde\alpha_{12}(N_1,1,x_3)=
2\pi m\left(\frac{1}{N_1N_2}-1\right),
\ee
which yields
\be
\label{equ:one_over_n_squared}
T_0=2\pi^2 \beta_G^2 \int_0^1\d m
\frac{2m}{N_1N_2}=\frac{2\pi^2\beta_G^2}{N_1N_2}.
\ee
In the Higgs phase, all the fields are massive, and the finite size
effects are exponentially suppressed.

\section{The relation of vortex tension to thermodynamics}
\label{sect:relation}

The behaviour of the vortex tension $T$ can be 
directly related to possible phase transitions in the system.
This is simply because a vortex carries a magnetic flux, and
a magnetic field contributes to the free energy. 
Let us discuss this in some more detail.

To begin with, note that 
a system with winding $m$ corresponds to the magnetic flux density
$B_z=2\pi m\beta_G^2/N_1N_2$. Thus, the free energy $F(B_z)$ of a 
system through which goes a non-vanishing magnetic flux, 
equals $F_m$ in \eq(\ref{equ:freeen}), with $m=a^2N_1N_2B_z/2\pi$. 
For symmetry reasons,  $F(B_z)=F(|B_z|)$. 
In the thermodynamical limit, $B_z$ can have any
real values. Hence the derivative of $F(B_z)$ becomes meaningful,
and we can calculate it from the vortex tension,
if we assume a repulsive
interaction between the vortices, as is observed at large~$x$.
A system with a finite number $m$ of vortices 
becomes namely infinitely dilute as the volume increases,
yielding $F_m=mF_1$, and thus
(we denote $f(B_z)=V^{-1}F(B_z)$, the free energy density
at a fixed flux density $B_z$),
\be
\left.\frac{\partial f(B_z)}{\partial |B_z|}\right|_{B_z=0}=
\frac{1}{V}\lim_{|B_z|\rightarrow 0}
\frac{F(B_z)-F(0)}{|B_z|}=
\frac{a^2N_1N_2}{2\pi V}(F_1-F_0)=
\frac{T}{2\pi},
\ee
where $V=a^3N_1N_2N_3$ is the volume of the system.

Let us denote by $g(H_z)$ the free energy density of a system in which 
the flux is allowed to fluctuate and an external field $H_z$ is applied:
\be
g(H_z)=-\frac{1}{V}\ln \int_{-\infty}^\infty \d B_z 
{\rm e}^{-V(f(B_z)-H_zB_z)}
\mathrel{\mathop{\longrightarrow}\limits_{V\rightarrow\infty}}
\min_{B_z}\left(f(B_z)-H_zB_z\right).
\label{Ltransf}
\ee
In the thermodynamical limit, $g(H_z)$ is thus given by the Legendre
transform of $f(B_z)$.

\begin{figure}
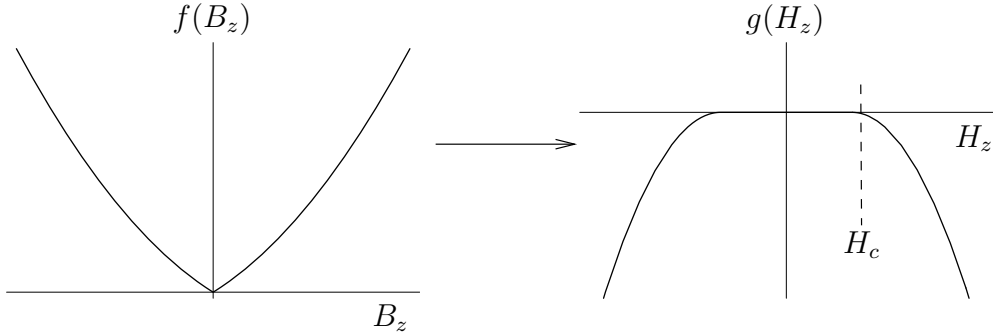

\input legendre.pstex_t

\vspace{-1cm}

\caption{
\label{fig:legendre}
If $T>0$, there is a cusp in the free energy density $f(B_z)$ at $B_z=0$.
Thus the function $g(H_z)$, defined in
\eq(\ref{Ltransf}), has a flat region at
$-H_c\le H_z\le H_c$.
}
\end{figure}

If $T>0$, the derivative of $f(B_z)$ is
discontinuous at $B_z=0$ (see Fig.~\ref{fig:legendre}). 
Its Legendre transform $g(H_z)$
will therefore have a flat region, i.e. $g(H_z)=g(0)$
when $|H_z|\le H_c\equiv T/2\pi$. This 
is precisely the Meissner effect, since 
$\langle B_z\rangle=-\partial g/\partial H_z=0$, if $|H_z|\le H_c$.
It also
implies that
\be
\label{equ:nonanal}
\left.\frac{\partial^2 g}{\partial H_z^2}\right|_{H_z=0}
\begin{array}{ll}
=0&, $ when$\quad T>0,\\
>0&, $ when$\quad T=0.
\end{array}
\ee
Thus, the values of $T$ are related to the analytic
structure (in particular, phase transitions)  of the free energy.

At $x=2$, we expect $T$ to behave continuously.
It is non-zero in the Higgs phase, but decreases
with increasing $y$, approaching zero at some value $y=y_c$.
At larger values of $y$, the tension stays zero.
If this turns out to be true, 
Eq.~(\ref{equ:nonanal}) shows that the free energy density $g$
is non-analytic at $y=y_c$.

\section{Simulations and results}
\label{sect:simulations}
The Monte Carlo simulations of the theory in \eq(\ref{equ:lattact}) at 
zero and non-zero values of~$m$ have been 
performed at the parameter value $x=2$, at $\beta_G=4$.
For this value of $\beta_G$, the correlation
lengths of massive modes like the scalar or the photon
in the Higgs phase are large in units of the lattice spacing $a$.
Indeed, for the values of $y$ considered here,
$-1.6 \le y \le 0.5$, 
even the shortest correlation length, which
is the scalar one, is always larger than about $3a$. 
Thus finite $a$ corrections to the measured 
quantities are expected to be small.   

The major part of the numerical work has been performed with the
help of Hybrid Monte Carlo (HMC) update algorithms, quite similar to those
used in lattice QCD simulations. HMC updates were 
implemented separately for the gauge fields and for  
the scalar fields. Within both HMC algorithms a leapfrog 
discretization based upon $8$ time steps into the molecular dynamics
time direction was chosen. Acceptance rates were appropriately
tuned. We measure our statistics in units of sweeps, where each
sweep consists of two completed trajectories, one in the gauge and one
in the scalar fields. The use of HMC updating schemes 
is favored by the fact that the usual checkerboard decomposition of degrees 
of freedom can be avoided and, therefore, the 
adaptation of the computer code 
to a parallel architecture is simplified. Most of our simulations 
have been executed on parallel computers. Some simulations have also been 
repeated on work stations with the use of different random number generation 
algorithms and therefore different random number sequences. These runs serve 
as a valuable cross-check, and no inconsistencies were detected.

We also experimented with over-relaxation  updates of the gauge fields
alone. As we failed to achieve a significant improvement in the statistical 
quality of the simulation, we will not report on the details of that
approach here.

\begin{figure}

\vspace{-3.0cm}

\hspace*{1cm}\epsfig{file=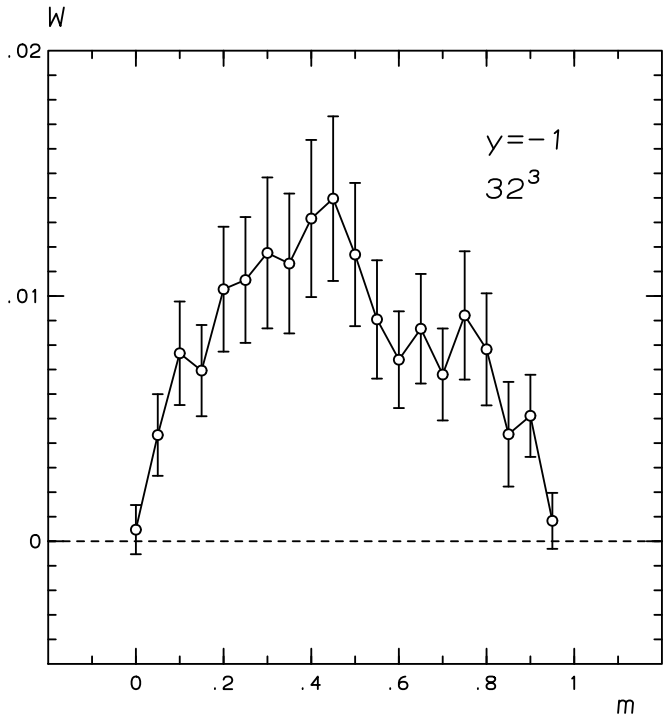,width=12cm}

\vspace{-3.5cm}

\caption{
\label{fig:fig1}
$W(m)$ on a $32^3$ lattice at $x=2$, $\beta_G=4$ and $y=-1$.}
\end{figure}

We simulate symmetric cubic lattices with volumes $N^3$.
Typical lattice sizes range from $N=16$
to $N=32$, which currently represents our largest system. The 
expectation value $W(m)$ in Eq.~(\ref{equ:double_of_m}) is evaluated
by means of Monte Carlo
simulations on a discrete set of $21$ values $m_i$
partitioning the interval $m=[0,1]$ into $20$ equally spaced subintervals.
The integral of \eq(\ref{equ:double_of_m}) was then calculated by the
trapezoidal rule and it was checked that systematic errors introduced by
the discretization are much smaller than statistical errors. A typical
Monte Carlo result for the quantity $W(m)$ 
is displayed in Fig.~\ref{fig:fig1}
for the $32^3$ lattice at $y=-1$. Statistical errors on each individual $m_i$ 
measurement can be as large as 40\%, as is expected for an 
energy difference calculation on such large systems. We employed
between $10000$ and $20000$ Monte Carlo sweeps on each of the
individual $m_i$ measurements. Statistical errors for the vortex
tension $T$ were then calculated using jackknife error calculation and 
statistical error propagation.

\begin{figure}

\vspace{-3.0cm}

\hspace*{1cm}\epsfig{file=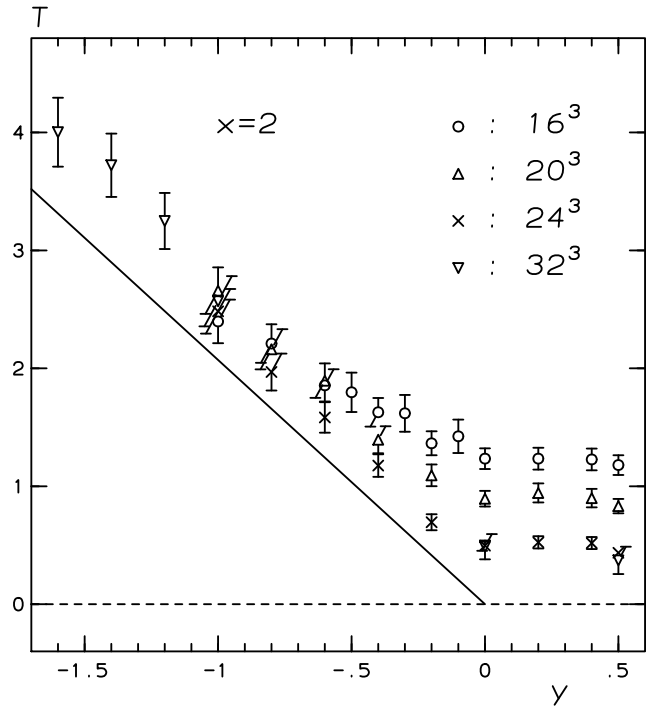,width=12cm}

\vspace{-3.5cm}

\caption{
\label{fig:fig2}
Lattice results for $T$ as a function of $y$ on $16^3$, $20^3$, $24^3$
and $32^3$ lattices at $x=2$ and $\beta_G=4$. The solid straight line is 
the mean-field result. Actually, only its slope is unambiguously 
determined.}
\end{figure}

In Fig. \ref{fig:fig2} we display our finite-volume data for the
vortex tension $T$ as a function of $y$ in the whole 
interval of $y$ considered
on a selected set of $16^3$, $20^3$, $24^3$ 
and $32^3$ lattices.
One identifies three different regions. For 
$y \le -1$ (Higgs phase), the vortex tension $T$ appears to be finite and 
finite size effects appear to be under control. Note that at $y=-1$, data
from several lattice sizes cluster around a common mean value.
For values of $y$ larger than about $y=0$ (Coulomb phase),
$T$ is non-vanishing in a finite system. 
However, it shows large finite size effects, which appear to be 
rather independent of $y$. There is also
a region of intermediate values of $y$ in which finite size effects
are present and presumably not of a simple form. In this paper we will 
concentrate on those values of $y$ that are clearly located in the Higgs
or the Coulomb phase. A study of the intermediate $y$ region would 
require finite size scaling methods beyond the scope of the present paper. 

\begin{figure}

\vspace{-2cm}

\hspace*{-1.4cm}\epsfig{file=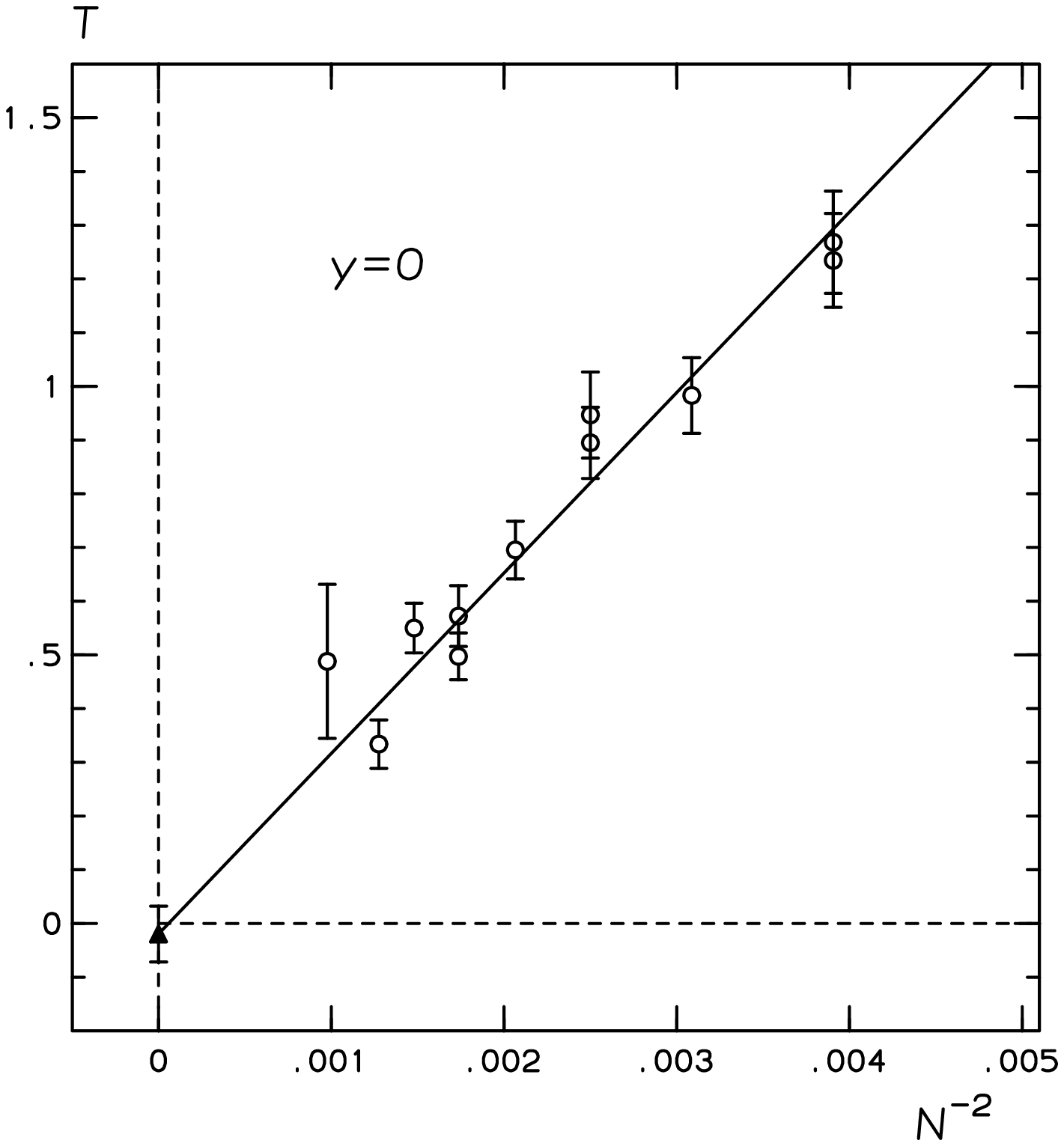,width=9cm}\hspace*{-1.5cm}%
\epsfig{file=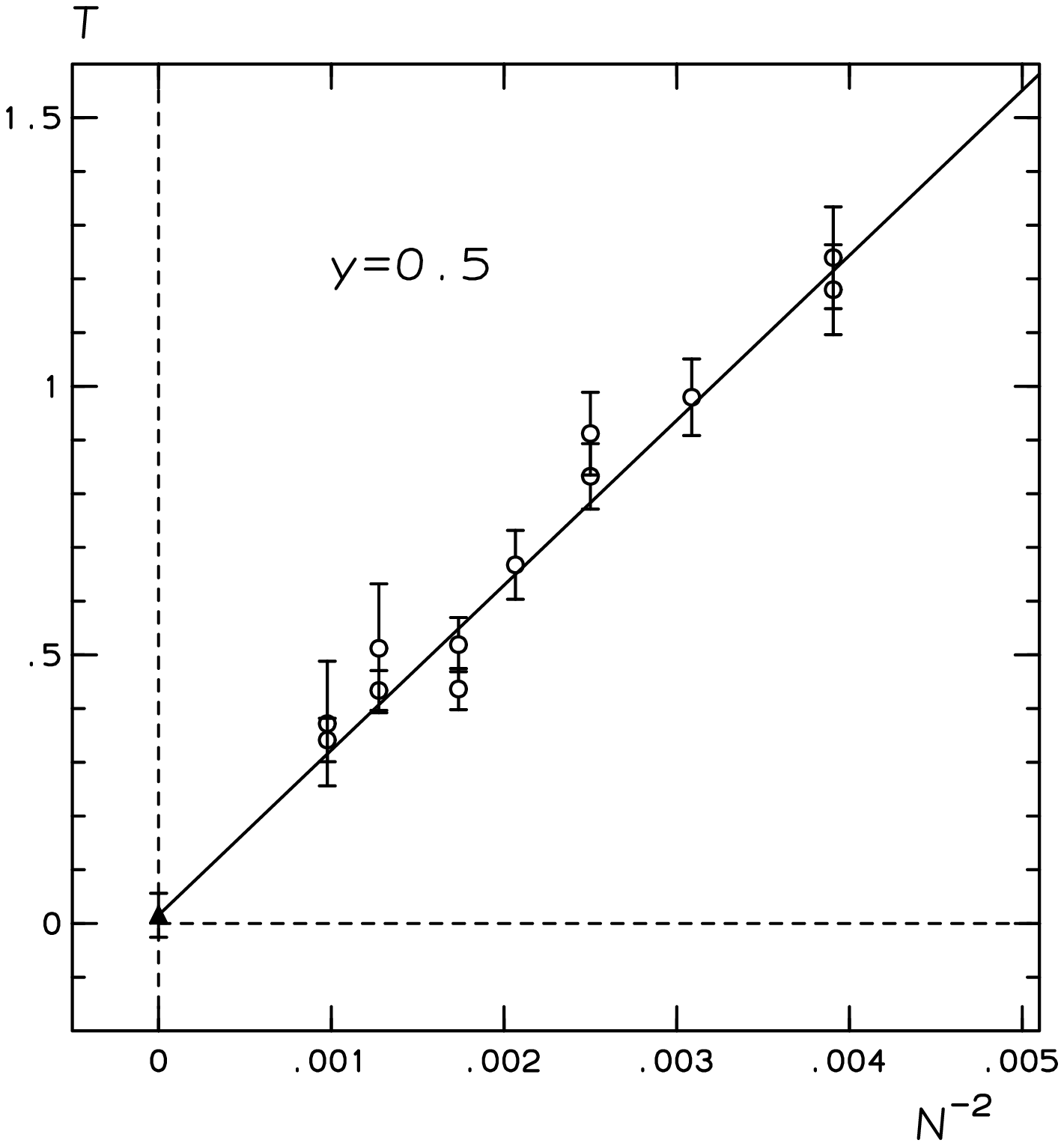,width=9cm}

\vspace{-2cm}

\caption{
\label{fig:fig3}
Finite size scaling data for $T$ at $x=2$, $\beta_G=4$, 
$y=0.0$ (left) and $y=0.5$ (right),  
as a function of $N^{-2}$. The solid line corresponds to a straight line fit 
as described in the text. The triangle corresponds 
to the infinite volume extrapolation.}
\end{figure}

For two values of $y$, $y=0$ and $y=0.5$, we investigated in detail 
the possible existence of a power-like finite size 
behavior $T \propto N^{-2}$, as is characteristic of 
the pure gauge theory (see 
Eq.~(\ref{equ:one_over_n_squared})). The observation of such a finite size 
behavior at the given values of $y$ within the Coulomb phase 
signals the homogeneity of the system under the response of an
applied flux. It is a strong argument in favor of the existence 
of a massless mode or alternatively, a vanishing vortex
tension $T$. In Fig. \ref{fig:fig3} we 
display our finite $N$ data for the vortex tension $T$ as a function of 
$N^{-2}$ at $y=0$ and $y=0.5$. The lattice sizes considered are 
$N^3=16^3,18^3,20^3,22^3,24^3,26^3,28^3$ and $32^3$ at $y=0$. The same 
lattices except $N^3=26^3$ have been considered at $y=0.5$.

\begin{figure}

\vspace{-3.0cm}

\hspace*{1cm}\epsfig{file=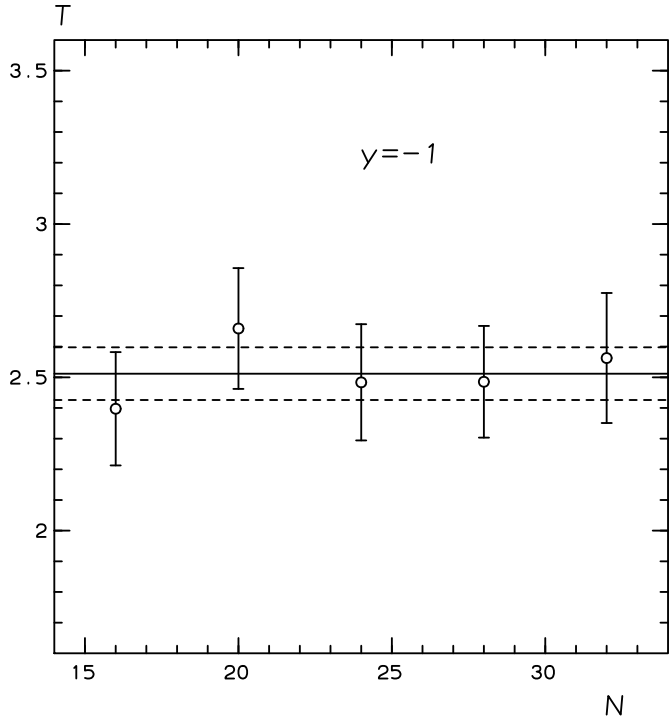,width=12cm}

\vspace{-3.5cm}

\caption{
\label{fig:fig5}
Finite size scaling data for $T$ at $y=-1$, $x=2$ and 
$\beta_G=4$, as a function of $N$. There are no visible finite size effects. 
The horizontal solid line corresponds to the infinite volume 
extrapolation $T=2.51(9)$, 
and the dashed lines indicate the error interval.}
\end{figure}

Both data sets 
have been independently fitted with the two-parameter form
\be
T(N)=R_0\frac{2\pi^2\beta_G^2}{N^2}+T(N=\infty). \label{equ:ansatz}
\ee
Deviations of the parameter $R_0$ from unity measure deviations 
from a behavior anticipated in pure gauge theory, and the parameter
$T(N=\infty)$ corresponds to our infinite volume extrapolation
of the vortex tension $T$. The
fit to all the data, as indicated by the solid straight lines
in Fig. \ref{fig:fig3}, gives 
\ba
\label{equ:R0Tfit1}
R_0 & = & 1.06(7),~~~T(N=\infty)=-0.019(32)~~~{\rm at}~~~y=0,~ \\
\label{equ:R0Tfit2}
R_0 & = & 0.97(6),~~~T(N=\infty)=+0.015(41)~~~{\rm at}~~~y=0.5.
\ea
We notice that $R_0$ agrees with unity within errorbars. It 
thus appears that the possible renormalization of the coefficient in front
of the $N^{-2}$ behavior induced by the presence of the scalar field 
is negligible.
Both fits exhibit reasonably small $\chi^2_\rmi{dof}$ values, namely
$\chi^2_\rmi{dof}=1.5$ at $y=0$ and $\chi^2_\rmi{dof}=1.4$ at $y=0.5$. 
Nevertheless, it can 
be noted that there are some data points which scatter. Simulations in the 
Coulomb phase turn out to be statistically rather demanding. From
Fig.~\ref{fig:fig5} we
observe that the Monte Carlo simulation
is better behaved (all errorbars cross the central value)
in the Higgs phase, which is natural since there is no
massless mode there.

One may alternatively fix the parameter value
$T(N=\infty)$ to zero in \eq(\ref{equ:ansatz}), and attempt to 
describe the data with
\be
T(N)={{\tilde R}_0}\frac{2\pi^2\beta_G^2}{N^\eta},
\ee
where now any deviation of the parameter $\eta$ from the value $2$  
signals departure from homogeneity caused by the
presence of a vortex. We find
\ba
\eta & = & 1.94(14)~~~{\rm at}~~~y=0,~ \\
\eta & = & 1.93(12)~~~{\rm at}~~~y=0.5,
\ea
and comparable values of the parameter ${{\tilde R}_0}$ as in the previous 
case.  These values clearly agree with the theoretical predictions for a 
Coulomb phase.

In the Higgs phase, at $y=-1.0$, the vortex tension $T$
was determined on $16^3$, $20^3$, $24^3$ and $32^3$ lattices, see
Fig. \ref{fig:fig5}. Within the statistical errors
the data are constant. A fit of the data to a constant value
has $\chi^2_\rmi{dof}=0.3$
and gives
\be
T(N=\infty)=2.51(9)~~~{\rm at}~~~y=-1.~
\ee
Comparing this value with the mean-field result $T_{\rm MF}=2.07$ from
Eq.~(\ref{equ:classT}),
we observe a $17$\% difference, i.e., the mean-field result is 
close to the measured value.
One source for the discrepancy is that our result was measured at
a finite lattice spacing~$a$, 
while the mean-field calculation was performed in the continuum.
One should also note that actually the mean-field calculation
only predicts the slope $\d T/\d y$ (see Fig.~\ref{fig:fig2}). 
Namely, the choice of the 
renormalization scale $\mu$ is arbitrary at the mean-field level. 
If we chose some other scale,
the value of $y$ would change by a constant term,
$y(\mu)=y(1)+(4-8x+8x^2)/(16\pi^2)\ln\mu$. 
In fact, changing from $\mu=1$ to $\mu=0.2$ would bring the results
in perfect numerical agreement. To remove this ambiguity,
corrections to the mean-field result would have to be computed up to
2-loop level.

Finally, we point out that since
the infinite volume extrapolation of the vortex free energy at the values
$y=0.0$ and $y=0.5$ is fully compatible with a vanishing vortex tension,
and at $y=-1$ with a non-vanishing value, the vortex tension
indeed behaves as an order parameter. 

\section{Conclusions}
\label{sect:conclusions}
We have shown by numerical simulations that the vortex tension $T$ defined
in Eq.~(\ref{equ:tension}) acts as an order parameter at large values of
$x$,
where the transition between the Higgs and the Coulomb phases
is continuous. The point $y_c$, at which $T$ vanishes, coincides with the
true transition point, since the free energy density $g$ was shown to be
non-analytic at $y_c$. 

At $y<-1$, the measured values of $T$ agree with the mean-field
result in \eq(\ref{equ:classT}) within errorbars, when the ambiguity of the
renormalization scale in the mean-field approximation is taken into
account.
At $y>0$, the system was observed to behave very much like 
pure gauge theory, i.e.~the change due to the presence of a
scalar field was very small.
Outside the vicinity of $y_c$, 
mean-field approximation thus seems to describe
vortices well in both phases of the theory even at $x=2$, where fluctuations
are large. However, we expect
that at $y\approx y_c$, there will be large deviations from the mean-field
behavior. The study of this interesting region requires
significantly more computational resources than used here.

We performed the simulations at a finite value of the lattice spacing $a$.
Although we believe that the vortex tension $T$ has a well-defined
continuum limit, we expect that there are ${\mathcal O}(a)$ errors, 
in particular in the Higgs phase. However, as of now, we have 
not removed them by a continuum extrapolation $a\to 0$.

In cosmology, the time evolution of the vortex network after a 
phase transition is typically treated at the mean-field level
\cite{ref:vincent}.
Our results give confidence to that approximation,
at least deep in the Higgs phase.
However, to see if modifications are needed to the picture
of defect formation in gauge theories,
a more detailed study is needed
near the transition temperature to find out the role of the fluctuations. 
For a complete picture of cosmic string formation, one would also
have to take the non-equilibrium effects into account.

Another interesting way to extend this first-principles numerical 
study would be to consider configurations with multiple vortices.
In the context of superconductors, it has been suggested 
and also numerically verified in simplified models that the
interactions between vortices give rise to new phases.
It is then theoretically interesting and computationally
demanding to explore, how the non-analytic structure
in the multiple vortex case is related to the phase
diagram at a vanishing magnetic flux density, Fig.~\ref{fig:phases}. 

\section*{Acknowledgements}

Most of the simulations were carried out with a Cray T3E at the Center 
for Scientific Computing, Finland. This work was
partly supported by the TMR network {\em Finite Temperature Phase
Transitions in Particle Physics}, EU contract no.\ FMRX-CT97-0122.

\bibliographystyle{plain}
\bibliography{free}
\end{document}